# Gold nanocrystal-mediated sliding of doublet DNA origami filaments

Maximilian J. Urban[1,2], Steffen Both[3], Chao Zhou[1,2,*], Anton Kuzyk[1,4],
Klas Lindfors[5], Thomas Weiss[3], and Na Liu[1,2,*]

[1]Max Planck Institute for Intelligent Systems, Heisenbergstrasse 3, D-70569 Stuttgart, Germany
[2]Kirchhoff Institute for Physics, Heidelberg University, Im Neuenheimer Feld 227, D-69120 Heidelberg, Germany
[3]4th Physics Institute and Stuttgart Research Center of Photonic Engineering, University of Stuttgart, 70569 Stuttgart, Germany
[4]Department of Neuroscience and Biomedical Engineering, Aalto University, School of Science, P.O. Box 12200, FI-00076 Aalto, Finland
[5]Department of Chemistry, University of Cologne, Luxemburger Straße 116, 50939 Köln, Germany
e-mail:laura.liu@is.mpg.de, czhou@is.mpg.de

**Sliding is one of the fundamental mechanical movements in machinery. In macroscopic systems, double-rack pinion machines employ gears to slide two linear tracks along opposite directions. In microscopic systems, kinesin-5 proteins crosslink and slide apart antiparallel microtubules, promoting spindle bipolarity and elongation during mitosis. Here we demonstrate an artificial nanoscopic analog, in which gold nanocrystals can mediate coordinated sliding of two antiparallel DNA origami filaments powered by DNA fuels. Stepwise and reversible sliding along opposite directions is *in situ* monitored and confirmed using fluorescence spectroscopy. A theoretical model including different energy transfer mechanisms is developed to understand the observed fluorescence dynamics. We further show that such sliding can also take place in the presence of multiple DNA side-locks that are introduced to inhibit the relative movements. Our work enriches the toolbox of DNA-based nanomachinery, taking one step further toward the vision of molecular nanofactories.**

## Introduction

A living cell is a miniature factory containing a variety of motor proteins, which can perform complicated tasks powered by chemical energy.[1,2] Such natural wonders developed through billions of years of evolution surpass any human-made nanomachines[3,4] in complexity and sophistication. One class of proteins is rotary motors,[5] for instance the $F_O F_1$-adenosine triphosphate synthase,[6,7] which synthesizes adenosine triphosphate from adenosine diphosphate and phosphate. There are also important linear motors,[8] such as kinesin, dynein, and myosin proteins, which can move in discrete steps along long polymer tracks, playing essential roles in intracellular transport, self-organization, and cell division.[2,9]

In DNA nanotechnology, there has been an everlasting pursuit to construct DNA-based artificial systems that mimic motor proteins for the vision of molecular nanofactories.[10-29] Despite the fact that a complete understanding of these biological motors remains a formidable challenge, exciting progress has already been witnessed toward this futuristic goal, for instance, the realizations of nanoscale rotary apparatus self-assembled using multilayer DNA origami,[27] a DNA walker that programmably collects nanoparticle cargos along an origami assembly line[16] and so forth.[30-34] Nevertheless, a particular type of controlled motion, namely, relative sliding, which myosin proteins execute on actin filaments for muscle contraction[35] or kinesin-5 proteins drive between microtubules for proper segregation of chromosomes during mitosis[36] has not yet been fully explored and attempted using DNA nanotechnology. Here, inspired by the intriguing sliding function of motor proteins, we construct an artificial nanoscopic analog, in which relative movements between doublet DNA origami filaments is mediated by coordinated nanoscale motions of gold nanocrystals (AuNCs) powered by DNA fuels.

## Results

**Design and characterization of the sliding system**

Figure 1a shows the schematic of our sliding system. The two 14-helix DNA origami filaments folded by a self-assembly process[37,38] are crosslinked through two AuNCs (10 nm in diameter) in an antiparallel fashion. To ensure a correct orientation between the two filaments and also allow for structural flexibility, the filaments are connected at the two ends using the scaffold strand. Ten rows of footholds (coded 1-10) are extended from each origami filament as shown in Fig. 1a (also see Supplementary Figs. 1, 2 and Supplementary Data 1). They are evenly separated by 7 nm to establish five states. In each row, there are three binding sites with identical footholds. The foothold rows are reversely positioned along the two filaments, whose polarities are defined using '+' and '-' as shown in Fig. 1a. The two AuNCs are bound in between the filaments with the same combination of four foothold rows, *i.e.*, two foothold rows from each filament. This takes the inspiration from the homotetramer structure of the kinesin-5 protein, which comprises four motor domains, two on each end to interact with the microtubule tracks.[39] To *in situ* optically monitor the sliding process, ATTO 550 (donor) and ATTO 647N (acceptor) are tethered at one end of the DNA origami structure. In the symmetric configuration as shown in Fig. 1a, transient binding of a short 8-base-pair (bp) segment brings the two fluorophores in close proximity, allowing for fluorescence resonance energy transfer (FRET).[40]

Fig. 1b schematically describes the AuNC-mediated sliding mechanism. The two AuNCs play a dual role. First, they define the system configuration by crosslinking and aligning the two origami filaments in antiparallel. Second, they serve as pinion gears, mediating stepwise sliding of the filaments relative to one another. Such coordinated motions can be activated in that the two AuNCs are bound to the same combination of foothold rows resulting from the opposite polarities of the doublet filaments. Consequently, the AuNCs can be driven simultaneously by the same sets of DNA fuels. Compared to our previous work on plasmonic walkers, [25, 26] here the length of the DNA strands on the AuNCs and the number of the footholds on the origami have to be reduced and specifically designed to enable relative sliding (see Supplementary Fig. 2). Initially, both of the AuNCs are bound to footholds 3, 4, 7, and 8 through DNA hybridization. The rest of the foothold rows are deactivated using respective blocking strands. To start sliding, blocking strands 4 and 8 as well as removal strands $\bar{2}$ and $\bar{6}$ are added simultaneously. The blocking strands help to detach the AuNCs from foothold rows 4 and 8 through toehold-mediated strand displacement reactions[41] and subsequently block these two rows. This eliminates back sliding of the AuNCs, thus imposing directionality. Simultaneously, removal strands $\bar{2}$ and $\bar{6}$ activate foothold rows 2 and 6 through toehold-mediated strand displacement reactions, allowing the AuNCs to bind. As a result, both AuNCs execute 7 nm movements to slide the filaments. This introduces an overall 14 nm displacement, which is twice the step size of each AuNC. Upon addition of corresponding DNA fuels for the next step, the AuNCs progressively slide the filaments following a similar aforementioned principle, introducing a 28 nm displacement.

Transmission electron microscopy (TEM) was carried out to examine the assembled AuNC-origami structures. An overview TEM image of the structures before sliding is shown in Fig. 2a, demonstrating a high yield of the DNA origami structures crosslinked by AuNCs (see also Supplementary Fig. 3 for DNA origami without AuNCs). The averaged TEM image (see Fig. 2a inset) reveals an excellent structural homogeneity of the symmetric configuration. Fig. 2b presents the TEM image of the structures after two sliding steps. The averaged TEM image (see Fig. 2b inset) shows an evident displacement between the two origami filaments (see also Supplementary Fig. 4), confirming successful sliding. The deviation of the displacement size from the nominal 28 nm is due to structural deformations on the TEM grid from the drying process.

Such AuNC-mediated sliding gives rise to dynamic distance variations between the two fluorophores, leading to FRET signal changes. To *in situ* optically monitor the sliding process, fluorescence signals of the sample were recorded using the time-scan function of a fluorescence spectrometer (Jasco-FP 8500) at two emission wavelengths of 578 nm (donor) and 663 nm (FRET) with an excitation wavelength of 530 nm. Fig. 3a illustrates the five states. The transitions between different states are powered by corresponding DNA fuels as discussed in Fig. 1b (also see Supplementary Table 1). The sliding process starts from the symmetric configuration, *i.e.*, state '0' (see Fig. 3a). Upon addition of the DNA fuels, the AuNCs execute one sliding step and the system reaches state '-1', introducing a 14 nm displacement between the two filaments. Both the FRET and donor signals experience clear intensity decreases (see Fig. 3b and Supplementary Note 1). The FRET intensity decrease indicates that the transient binding between the fluorophore-bearing strands is reduced resulting from filament sliding, but it may still take place to some extent due to the flexibility of the single-stranded DNA linkers (see Supplementary Table 2). Meanwhile, the donor signal is slightly quenched as the right AuNC gradually approaches the donor fluorophore through sliding. When the AuNCs execute one more sliding step to state '-2' for an overall 28 nm displacement, the FRET and donor signals show further decreases. To demonstrate reverse sliding, the system is driven back to state '0' by sliding along the opposite direction. As shown in Fig. 3b, both the FRET and donor signals display two clear steps, approximately returning to their respective levels at state '0'. The observed signal degradations are likely due to sample imperfections. Subsequently, the AuNCs continue to slide the filaments, reaching states '1' and '2', respectively. The FRET signal exhibits two decreasing steps, whereas the donor signal increases successively in two steps. The former is due to the enlarged distance between the two fluorophores, similar to the previous cases at states '-1' and '-2'. The latter is mainly due to the considerable suppression of the donor quenching[42] in addition to reduced FRET, given the increasing separations between the donor and the right AuNC. Afterwards, the system carries out another reverse sliding back to state '0', accompanied with two individual steps in FRET and donor signals, respectively. In this regard, the sliding process can be successfully monitored by optical spectroscopy in real time. Particularly, the donor signal tracking offers an extra benefit to optically resolve the relative sliding directions, afforded by the quenching effects between the AuNC and the donor fluorophore (see Supplementary Fig. 5).[42]

**Theoretical calculation of the optical response changes**

To elucidate the underlying physics, we have developed a theoretical model to quantitatively describe the observed optical response changes. There are two main energy transfer mechanisms involved in our sliding system. One is associated with the electromagnetic interaction between the fluorophores and the nearby AuNC (the influence from the farther AuNC can be neglected). The AuNC exhibits a plasmonic resonance that modifies the radiative and nonradiative decay channels for the emission of photons from the fluorophores (see Supplementary Fig. 10).[43-45] For the AuNC as small as 10 nm, its absorption loss dominates, resulting in quenching of the fluorophores that are placed in close proximity (see Supplementary Figs. 5 and 11).[46] The other is associated with the electromagnetic interaction between the two fluorophores. When the donor and the acceptor approach one another, FRET starts to take place. Both mechanisms contribute to the fluorescence signals of the donor and the acceptor, which are proportional to the respective fluorescence rates, *i.e.*, the number of photons emitted per second. The fluorescence rates of the donor and the acceptor are

$$\gamma_{\text{fl}}^{\text{D}} = q^{\text{D}} \gamma_{\text{exc}}^{\text{D}} \text{ and } \gamma_{\text{fl}}^{\text{A}} = q^{\text{A}} q_{\text{FRET}} \gamma_{\text{exc}}^{\text{D}}, \qquad (1)$$

respectively, where $\gamma_{\text{exc}}^{\text{D}}$ is the excitation rate of the donor, *i.e.*, the number of photons absorbed per second; $q_{\text{FRET}}$ is the FRET efficiency, defined as the number of excitations transferred from the donor to the acceptor over the total number of donor excitations; $q^D$ and $q^A$ are the quantum yields of the donor and the acceptor, respectively, *i.e.*, the ratio between the number of radiated photons and the number of total excitations by the corresponding fluorophore (see Supplementary Fig. 10 for details on the fluorophores).

Subsequently, the FRET efficiency can be written as

$$q_{\text{FRET}} = \frac{\gamma_{\text{FRET}}}{\gamma_r^D + \gamma_{\text{abs}}^D + \gamma_{\text{nr}}^D + \gamma_{\text{FRET}}} . \tag{2}$$

The quantum yields of the donor and the acceptor are

$$q^D = \frac{\gamma_r^D}{\gamma_r^D + \gamma_{\text{abs}}^D + \gamma_{\text{nr}}^D + \gamma_{\text{FRET}}} \text{ and } q^A = \frac{\gamma_r^A}{\gamma_r^A + \gamma_{\text{abs}}^A + \gamma_{\text{nr}}^A} , \tag{3}$$

respectively, where $\gamma_{\text{FRET}}$ is the FRET rate that describes the energy transfer between the two fluorophores; $\gamma_r^D$ and $\gamma_r^A$ are the corresponding radiative decay rates of the donor and the acceptor in the presence of the AuNC, respectively; $\gamma_{\text{nr}}^D$ and $\gamma_{\text{nr}}^A$ are the intrinsic nonradiative decay rates of the donor and the acceptor, respectively, which are assumed not to be modified by the AuNC; $\gamma_{\text{abs}}^D$ and $\gamma_{\text{abs}}^A$ are the rates of energy transfer from the donor and the acceptor to the AuNC, respectively. The AuNC also increases $\gamma_{\text{exc}}^D$ due to the local near-field enhancement of the incident light.

The different rates in Eqs. (1), (2), and (3) can be derived from numerical simulations. In brief, $\gamma_r^D, \gamma_r^A, \gamma_{\text{abs}}^D$, and $\gamma_{\text{abs}}^A$ are obtained from the electromagnetic near fields excited by point dipoles at the fluorophore positions, while the enhancement of $\gamma_{\text{exc}}^D$ is acquired from the near fields at plane wave incidence. The presence of the AuNC also influences $\gamma_{\text{FRET}}$.[47] This effect, however, is negligible for all the donor-acceptor configurations in our system. Instead, $\gamma_{\text{FRET}}$ was calculated using the analytical FRET equation,[48] *i.e.*, $\gamma_{\text{FRET}} \propto R_0^6/R^6$, where $R$ is the distance between the fluorophores and $R_0$ is the Förster radius (see Supplementary Fig.

12 and Supplementary Table 2 for details on positions and distances). From the FRET efficiency, the quantum yields and the excitation enhancement, the FRET and donor fluorescence intensities can be calculated. The calculated FRET and donor fluorescence signals at the emission peak wavelengths are presented by the red and black curves in Fig. 3c, respectively. Our theoretical results exhibit an overall agreement with the experimental results in Fig. 3b. Details on the theoretical model and related derivations can be found in Supplementary Note 2.

**Sliding in the presence of DNA side-locks**

To provide insights into the AuNC-mediated sliding under inhibition, four DNA side-locks (i-iv) are introduced at the two ends of the DNA origami structure to restrain relative movements as shown in Fig. 4a. The two fluorophores are omitted in this figure for clarity. Locks i and iii are designed to be identical. So are locks ii and iv. At each end, the two DNA locks (i and ii or iii and iv) differ from one another by a 12-nucleotide (nt)-long locking sequence (purple or blue) as well as a distinct toehold segment (green or red) (see Supplementary Fig. 2). Each DNA lock contains two arms. For instance, for the side-lock iv (see Fig. 4b), one arm possesses a 31-bp DNA segment (black) with a 12-nt-long locking sequence (purple). The other arm possesses its complementary locking sequence (purple) and a toehold segment (green). Such a DNA side-lock can be locked or unlocked through toehold-mediated strand displacement reactions upon addition of corresponding DNA fuels as shown in Fig. 4b.

Three sample copies (A, B, and C) have been prepared to optically characterize the sliding kinetics in dependence on the DNA side-lock number using *in situ* fluorescence spectroscopy (see Supplementary Table 3). The samples start at state '0'. As indicated by the grey arrow in Fig. 4c, upon addition of the unlocking strands for locks ii and iv in sample B, the FRET signal experiences a slight decrease (blue line) in that the two fluorophores gain more spatial flexibility with only two remaining locks (*i.e*., i and iii), when compared to the case in sample A (red line), in which all the four locks are locked. This becomes more obvious for sample C (black line), which exhibits a larger FRET signal decrease in comparison to sample B, as all the four locks are unlocked. Subsequently, these samples with four (sample A), two (sample B), and zero (sample C) locks undergo stepwise sliding. When transitioning to state '1', samples A, B, and C exhibit the smallest, intermediate, and largest FRET signal decreases accordingly. This demonstrates clear inhibitions of the relative movements from the DNA locks as well as reveals an evident dependence of the sliding efficiency on the side-lock number (see Supplementary Fig. 6 for unlocking during the sliding process from '0' to '1'). Subsequently, the samples transit from state '1' to state '2' for further sliding upon addition of the corresponding DNA fuel strands. In addition, the donor signal tracking nicely substantiates the observations from the FRET signal tracking as shown in Fig. 4c.

To further examine the AuNC-mediated sliding behavior under inhibitions, detailed TEM structural analysis of different samples has been carried out after two sliding steps from state '0' in dependence on the presence of the DNA side-locks as well as on the AuNC number. Fig. 4d presents the histograms of the end-to-end distance (s) between the filaments after two sliding steps. The end-to-end distance instead of the relative displacement is utilized as the structural quantity for easier analysis (see Supplementary Figs. 7 and 8). System (1) with two AuNCs and without DNA locks presents a single maximum as shown in Fig. 4d. System (2) with one AuNC and without DNA locks exhibits a very similar trend yet with a broader distribution. This reveals in principle one single AuNC can also mediate relative movements, elucidating the robustness of the sliding mechanism. Two maxima are observed for system (3), which possesses two AuNCs and four locks. The occurrence of the new maximum at a shorter end-to-end distance corroborates the sliding inhibitions from the DNA locks observed in Fig. 4c. Apparently, system (4) with one AuNC and four locks contains the highest number of structures that remain locked and unable to slide among all the cases. This set of experiments exemplifies several important insights. First, one AuNC can work alone to slide apart the doublet DNA origami filaments, whereas two AuNCs can work together in a cooperative manner. Second, two AuNCs possess much higher sliding efficiency to break four DNA side-locks for enabling relative movements than one single AuNC [see systems (3) and (4)]. It is notable that the two AuNCs in system (3) [or even one AuNC in system (4)] can overcome the inhibitions from four DNA side-locks to mediate the sliding activities. Third, the averaged TEM images of the locked structures with one or two AuNCs manifest evident filament bending due to the presence of the four side-locks as shown in Fig. 4d, especially when compared to the unlocked structures in Fig. 2a. This adds further insights into the influence from the DNA side-locks on the structural conformations. The related kinetic and mechanical behaviors of the system certainly deserve further detailed investigations.[49-51]

**Discussion**

One of the exciting objectives in the field of DNA nanotechnology is to accomplish advanced artificial nanofactories, in which the spatial arrangements of different components and most dynamic behavior are enabled by DNA.[52] The first milestone is to learn and mimic how living organisms, for instance, motor proteins work and function so that one can gain a lot of insights into how DNA machines have to be built on the nanoscale. The second milestone is to master over coordination and communication among multiple DNA machines for efficient and productive capabilities.

Right now, the endeavors towards this ambitious yet extremely exciting goal are at the very beginning. We are still standing at a critical point of constructing individual bio-inspired DNA machines, which can perform very basic mechanical motions including rotating, walking, sliding, and so forth. Our studies of the artificial nanoscopic sliding systems are undoubtedly a valuable asset to enrich the tool box of DNA-based nano-machinery. Interesting follow-up research directions could be exploring other energy inputs including ATP hydrolysis or light for achieving system operation at higher rates and efficiencies, once again taking the inspiration from natural biological machines. Slowly gaining momentum, the realization of a variety of artificial DNA devices will envisage many fruitful outcomes, which in turn will advance DNA nanotechnology to a new dimension of functional potential. By then, we believe the vision of DNA-based molecular nanofactories will become quite achievable.

## Methods

**Design and preparation of the DNA origami filaments**

DNA scaffold strands (p8064) were purchased from Tilibit. All other DNA strands were purchased from Sigma-Aldrich or Eurofins (high-performance liquid chromatography purification for the thiol-modified and dye modified DNA and reverse-phase cartridge purification for the staple strands, capture strands, blocking strands, and removal strands).

The DNA origami structure was designed using caDNAno software. The DNA origami filaments consist of 2×14 helices arranged in a "honeycomb" lattice. To prevent the aggregation of the DNA origami, six thymine bases were added to the respective staple strands at the edge of the origami (design and sequence details can be found in Supplementary Figs. 1, 2 and Supplementary Data 1). The DNA origami structures were prepared by mixing 5 nM of the scaffold strands with 10 times of the staple strands and the respective capture strands in a buffer containing 0.5×TE ( TRIS, EDTA, pH=8), 20 mM $MgCl_2$, and 5 mM NaCl. The mixture was then annealed as follows: 85°C for 5 min; from 65°C to 61°C, 1°C /5 min; from 60°C to 51°C, 1°C /60 min; from 51°C to 38°C, 1°C /20 min; from 37°C to 26°C, 1°C /10 min; held at 25°C. The DNA origami structures were purified by PEG precipitation to remove the excess staple strands.

**Synthesis of the AuNCs**

AuNCs (10 nm) were synthesized using a two-step method. A 1.25 mL $HAuCl_4$ solution (0.2%, w/v) was diluted in 25 mL double-distilled water and heated to boiling. A 1 mL sodium citrate solution (1%, w/v; containing 0.05% citric acid) was added to the flask under vigorous stirring. The solution in the flask was kept boiling for 5 min under stirring and then cooled at room temperature.

**Surface modification of the AuNCs with BSPP**

Bis(p-sulfonatophenyl)phenylphosphine dihydrate dipotassium salt (BSPP) (15 mg) was added to the Au colloidal solution (20 mL, OD ~ 1) and the mixture was shaken overnight at room temperature. Sodium chloride (solid) was added slowly to this mixture solution, while stirring until the solution color was changed from deep burgundy to light purple. The resulting mixture was centrifuged at 500 rcf for 30 min and the supernatant was removed with a pipette. AuNCs were then resuspended in a 1 mL BSPP solution (2.5 mM). Upon mixing with 1 mL methanol, the mixture was centrifuged again at 500 rcf for 30 min. The supernatant was removed and the AuNCs were resuspended in a 1 mL BSPP solution (2.5 mM). The concentration of the AuNCs was estimated according to the optical absorption at 520 nm.

**Preparation of AuNC-DNA conjugates**

AuNC DNA conjugation was done according to Kuzyk *et al*. with minor modifications.[53] The disulfide bond in the thiol-modified oligonucleotides was reduced using tris(2-carboxyethyl)phosphine (TCEP) (100 mM, 1 h) in water. Thiol-modified oligonucleotides and BSPP modified AuNCs were then incubated at a molar ratio of DNA to AuNC of 300:1 in a 0.5×TBE buffer solution for 20 h at room temperature. The concentration of NaCl was slowly increased to 300 mM in the subsequent 20 h in order to increase the density of thiolated DNA on AuNCs. AuNC-DNA conjugates were then washed using a 0.5×TBE (tris-(hydroxymethyl)-aminomethan, borate, ethylenediaminetetraacetic acid) buffer solution in 100 kDa (MWCO) centrifuge filters to remove the free oligonucleotides. The concentration of the AuNC-DNA conjugates was estimated according to the optical absorption at 520 nm. Freshly prepared, fully coated AuNCs do not precipitate in a 0.5×TBE 10 mM $MgCl_2$ buffer.

**Self-assembly of the AuNCs on DNA origami**

First, 10 times excess of the blocking strands 1, 2, 5, 6, 9, and 10 were added to the purified DNA origami and incubated at room temperature for 0.5 h to block the footholds 1, 2, 5, 6, 9, and 10 (attachment of two AuNCs at positions 3, 4 and 7, 8). For the attachment of a single AuNC, blocking strands 1, 2, 3, 4, 7, 8, 9, and 10 were used. The spin-filtered AuNCs were added to the DNA origami structures in an excess of 10 AuNCs per binding site on the DNA origami structure. The mixture was incubated on a shaker for over 24 h at 23°C. An agarose gel purification step (0.5% agarose gel in a 0.5×TBE buffer with 11 mM $MgCl_2$) was used to purify the successfully assembled product.

**TEM characterization**

The DNA origami structures were imaged using Philips CM 200 TEM operating at 200 kV. For imaging, the DNA origami structures were deposited on freshly glow-discharged carbon/formvar TEM grids. The TEM grids were treated with a uranyl formate solution (0.75%) for negative staining of the DNA structures. Uranyl formate for negative TEM staining was purchased from Polysciences, Inc.. Average TEM images were obtained using EMAN2 software.[54]

**Fluorescence spectroscopy**

Fluorescence spectra were measured using a Jasco-FP8500 Fluorescence Spectrometer with a quartz SUPRASIL ultra-micro cuvette (path length, 10 mm). All measurements were carried out at room temperature in a buffer after agarose gel purification (0.5×TBE buffer with 11 mM $MgCl_2$, pH = 8). For the *in situ* fluorescence measurements, a 120 µL solution containing

~3 nM of the structures at the initial configuration was used. The fluorescence emissions at 578 nm and 663 nm were monitored using the time-scan acquisition mode and a data pitch of 1 s. The excitation wavelength was 530 nm. Respective blocking and removal strands were added to enable a programmed sliding.

**Data availability**

The data that support the plots within this paper and other findings of this study are available from the corresponding author upon reasonable request.

# References

1. Vale, R. D. The molecular motor toolbox for intracellular transport. *Cell* **112**, 467-480 (2003).
2. van den Heuvel, M. G. L. & Dekker, C. Motor proteins at work for nanotechnology. *Science* **317**, 333-336 (2007).
3. Browne, W. R. & Feringa, B. L. Making molecular machines work. *Nat. Nanotech.* **1**, 25-35 (2006).
4. Erbas-Cakmak, S., Leigh, D. A., McTernan, C. T. & Nussbaumer, A. L. Artificial molecular machines. *Chem. Rev.* **115**, 10081-10206 (2015).
5. Fillingame, R. H. Protein structure - molecular rotary motors. *Science* **286**, 1687-1688 (1999).
6. Stock, D., Leslie, A. G. W. & Walker, J. E. Molecular architecture of the rotary motor in ATP synthase. *Science* **286**, 1700-1705 (1999).
7. Sambongi, Y. *et al.* Mechanical rotation of the c subunit oligomer in ATP synthase ($F_0F_1$): Direct observation. *Science* **286**, 1722-1724 (1999).
8. Woehlke, G. & Schliwa, M. Walking on two heads: The many talents of kinesin. *Nat. Rev. Mol. Cell Biol.* **1**, 50-58 (2000).
9. Nedelec, F. J., Surrey, T., Maggs, A. C. & Leibler, S. Self-organization of microtubules and motors. *Nature* **389**, 305-308 (1997).
10. Sherman, W. B. & Seeman, N. C. A precisely controlled DNA biped walking device. *Nano Lett.* **4**, 1801-1801 (2004).
11. Shin, J. S. & Pierce, N. A. A synthetic DNA walker for molecular transport. *J. Am. Chem. Soc.* **126**, 10834-10835 (2004).
12. Yin, P., Yan, H., Daniell, X. G., Turberfield, A. J. & Reif, J. H. A unidirectional DNA walker that moves autonomously along a track. *Angew. Chem., Int. Ed.* **43**, 4906-4911 (2004).
13. Tian, Y., He, Y., Chen, Y., Yin, P. & Mao, C. D. Molecular devices - a DNAzyme that walks processively and autonomously along a one-dimensional track. *Angew. Chem., Int. Ed.* **44**, 4355-4358 (2005).
14. Green, S. J., Bath, J. & Turberfield, A. J. Coordinated chemomechanical cycles: A mechanism for autonomous molecular motion. *Phys. Rev. Lett.* **101**, 238101 (2008).
15. Omabegho, T., Sha, R. & Seeman, N. C. A bipedal DNA Brownian motor with coordinated legs. *Science* **324**, 67-71 (2009).
16. Gu, H. Z., Chao, J., Xiao, S. J. & Seeman, N. C. A proximity-based programmable DNA nanoscale assembly line. *Nature* **465**, 202-205 (2010).
17. Lund, K. *et al.* Molecular robots guided by prescriptive landscapes. *Nature* **465**, 206-210 (2010).
18. von Delius, M. & Leigh, D. A. Walking molecules. *Chem. Soc. Rev.* **40**, 3656-3676 (2011).
19. Wickham, S. F. J. *et al.* A DNA-based molecular motor that can navigate a network of tracks. *Nat. Nanotech.* **7**, 169-173 (2012).
20. Cha, T. G. *et al.* A synthetic DNA motor that transports nanoparticles along carbon nanotubes. *Nat. Nanotech.* **9**, 39-43 (2014).

21. Kuzyk, A. *et al.* Reconfigurable 3D plasmonic metamolecules. *Nat. Mater.* **13**, 862-866 (2014).
22. Gerling, T., Wagenbauer, K. F., Neuner, A. M. & Dietz, H. Dynamic DNA devices and assemblies formed by shape-complementary, non-base pairing 3D components. *Science* **347**, 1446-1452 (2015).
23. Pan, J., Li, F., Cha, T. G., Chen, H. & Choi, J. H. Recent progress on DNA based walkers. *Curr. Opin. Biotechnol.* **34**, 56-64 (2015).
24. Marras, A. E., Zhou, L. F., Su, H. J. & Castro, C. E. Programmable motion of DNA origami mechanisms. *Proc. Natl. Acad. Sci.* **112**, 713-718 (2015).
25. Zhou, C., Duan, X. & Liu, N. A plasmonic nanorod that walks on DNA origami. *Nat. Commun.* **6**, 8102 (2015).
26. Urban, M. J., Zhou, C., Duan, X. Y. & Liu, N. Optically resolving the dynamic walking of a plasmonic walker couple. *Nano Lett.* **15**, 8392-8396 (2015).
27. Ketterer, P., Willner, E. M. & Dietz, H. Nanoscale rotary apparatus formed from tight-fitting 3D DNA components. *Sci. Adv.* **2**, e1501209 (2016).
28. Kuzyk, A. *et al.* A light-driven three-dimensional plasmonic nanosystem that translates molecular motion into reversible chiroptical function. *Nat. Commun.* **7**, 10591 (2016).
29. Kuzyk, A., Urban, M. J., Idili, A., Ricci, F. & Liu, N. Selective control of reconfigurable chiral plasmonic metamolecules. *Sci. Adv.* **3**, e1602803 (2017).
30. Derr, N. D. *et al.* Tug-of-war in motor protein ensembles revealed with a programmable DNA origami scaffold. *Science* **338**, 662-665 (2012).
31. Wollman, A. J. M., Sanchez-Cano, C., Carstairs, H. M. J., Cross, R. A. & Turberfield, A. J. Transport and self-organization across different length scales powered by motor proteins and programmed by DNA. *Nat. Nanotech.* **9**, 44-47 (2014).
32. Sato, Y., Hiratsuka, Y., Kawamata, I., Murata, S. & Nomura, S.-i. M. Micrometer-sized molecular robot changes its shape in response to signal molecules. *Science Robotics* **2**, eaal3735 (2017).
33. Thubagere, A. J. *et al.* A cargo-sorting DNA robot. *Science* **357**, eaan6558 (2017).
34. List, J., Falgenhauer, E., Kopperger, E., Pardatscher, G. & Simmel, F. C. Long-range movement of large mechanically interlocked DNA nanostructures. *Nat. Commun.* **7**, 12414 (2016).
35. Cooper, G. M. *The cell: A molecular approach* Ch. Actin, Myosin, and Cell Movement, (Sinauer Associates, Sunderland (MA), 2000).
36. Kapitein, L. C. *et al.* The bipolar mitotic kinesin Eg5 moves on both microtubules that it crosslinks. *Nature* **435**, 114-118 (2005).
37. Rothemund, P. W. K. Folding DNA to create nanoscale shapes and patterns. *Nature* **440**, 297-302 (2006).
38. Douglas, S. M. *et al.* Self-assembly of DNA into nanoscale three-dimensional shapes. *Nature* **459**, 414-418 (2009).
39. Lodish, H. *et al. Molecular cell biology*. 8th edn, (W. H. Freeman, New York, 2016).
40. Clegg, R. M. Fluorescence resonance energy transfer. *Curr. Opin. Biotechnol.* **6**, 103-110 (1995).
41. Zhang, D. Y. & Seelig, G. Dynamic DNA nanotechnology using strand-displacement reactions. *Nat. Chem.* **3**, 103-113 (2011).
42. Dulkeith, E. *et al.* Fluorescence quenching of dye molecules near gold nanoparticles: Radiative and nonradiative effects. *Phys. Rev. Lett.* **89**, 203002 (2002).
43. Sauvan, C., Hugonin, J. P., Maksymov, I. S. & Lalanne, P. Theory of the spontaneous optical emission of nanosize photonic and plasmon resonators. *Phys. Rev. Lett.* **110**, 237401 (2013).
44. Akselrod, G. M. *et al.* Probing the mechanisms of large purcell enhancement in plasmonic nanoantennas. *Nat. Photon.* **8**, 835-840 (2014).
45. Muljarov, E. A. & Langbein, W. Exact mode volume and purcell factor of open optical systems. *Phys. Rev. B.* **94**, 235438 (2016).
46. Acuna, G. P. *et al.* Distance dependence of single-fluorophore quenching by gold nanoparticles studied on DNA origami. *ACS Nano* **6**, 3189-3195 (2012).

47. Marocico, C. A., Zhang, X. & Bradley, A. L. A theoretical investigation of the influence of gold nanosphere size on the decay and energy transfer rates and efficiencies of quantum emitters. *J. Chem. Phys.* **144**, 024108 (2016).
48. Novotny, L. & Hecht, B. *Principles of nano-optics*. 2nd edn, (Cambridge University Press, Cambridge, 2012).
49. Zhang, D. Y. & Winfree, E. Control of DNA strand displacement kinetics using toehold exchange. *J. Am. Chem. Soc.* **131**, 17303-17314 (2009).
50. Srinivas, N. *et al.* On the biophysics and kinetics of toehold-mediated DNA strand displacement. *Nucleic Acids Res.* **41**, 10641-10658 (2013).
51. Zhang, J. X. *et al.* Predicting DNA hybridization kinetics from sequence. *Nat. Chem.* **10**, 91-98 (2018).
52. Simmel, F. C. DNA-based assembly lines and nanofactories. *Curr. Opin. Biotechnol.* **23**, 516-521 (2012).
53. Kuzyk, A. *et al.* DNA-based self-assembly of chiral plasmonic nanostructures with tailored optical response. *Nature* **483**, 311-314 (2012).
54. Tang, G. *et al.* EMAN2: An extensible image processing suite for electron microscopy. *J. Struct. Biol.* **157**, 38-46 (2007).

**Acknowledgements**

This project was supported by the Sofja Kovalevskaja grant from the Alexander von Humboldt-Foundation, the Volkswagen foundation, and the European Research Council (ERC Dynamic Nano) grant. M. U. acknowledges the financial support by the Carl-Zeiss-Stiftung. A.K. was supported by the Academy of Finland (grant 308992). We thank Marion Kelsch for assistance with TEM. TEM images were collected at the Stuttgart Center for Electron Microscopy. We thank X. Shen for synthesis of the gold nanocrystals.

**Author Contributions:** M.U., C.Z, and N.L. conceived the project. M.U. and C.Z. designed the DNA origami nanostructures. M.U. performed all the experiments. S.B. and T.W. carried out the theoretical calculations. A.K. and K. L. made helpful suggestions. M.U. S.B. and N.L wrote the manuscript. All authors discussed the results and commented on the manuscript.

**Competing interests:** The authors declare no competing interests.

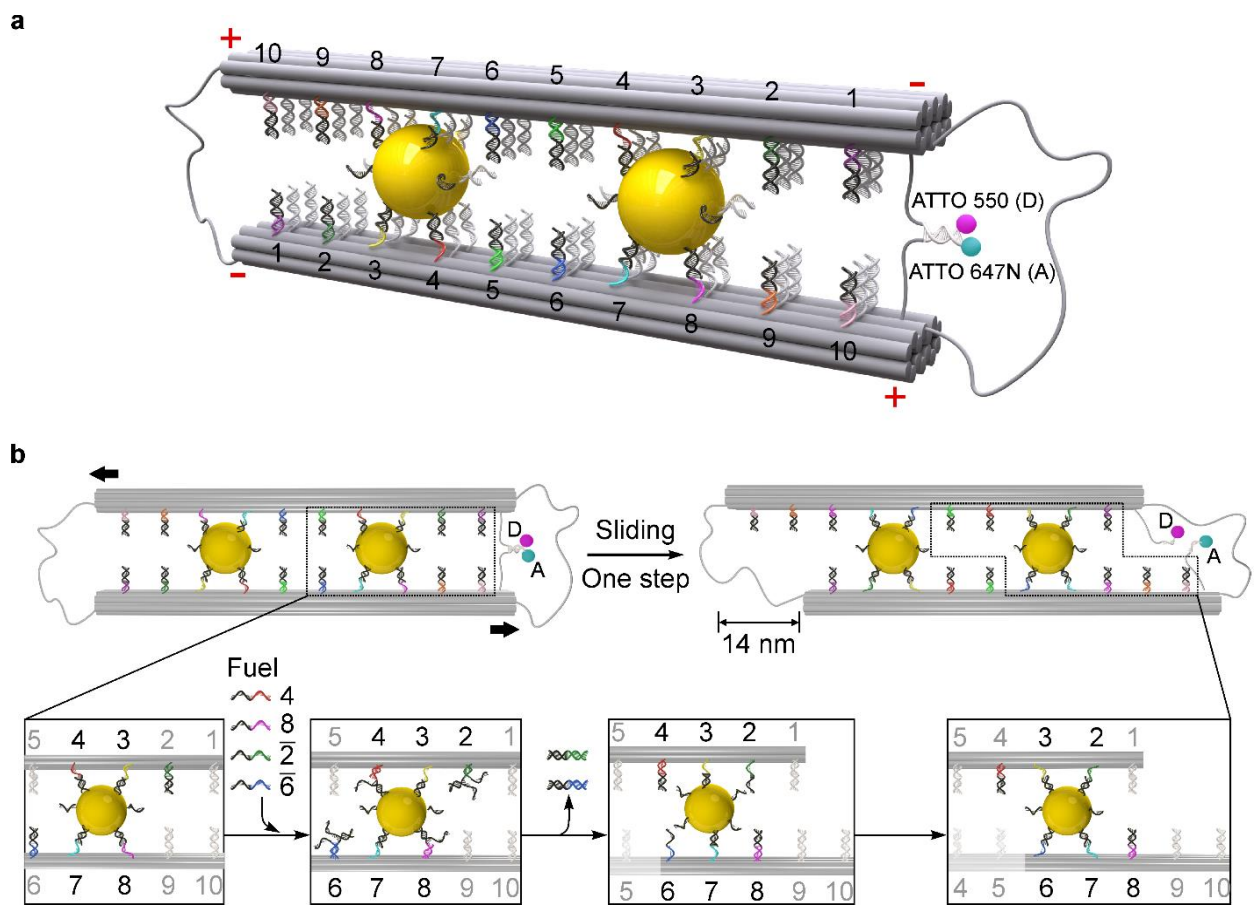

**Figure 1 | Schematic of the sliding system and working principle. a**, Two gold nanocrystals (AuNCs, 10 nm) crosslink two DNA origami filaments in antiparallel. Each filament comprises ten rows of footholds (coded 1-10) with 7 nm spacing. The foothold rows are reversely positioned along the two filaments, which possess opposite polarities as indicated using '+' and '-'. The filaments are connected by the scaffold strand to ensure a correct conformation as well as to enable structural flexibility. A pair of fluorophores (donor: ATTO 550 and acceptor ATTO 647N) are tethered at one end of the origami to allow for *in situ* optically monitoring the sliding dynamics via fluorescence resonance energy transfer (FRET). D and A represent donor and acceptor, respectively. **b**, Upon addition of blocking strands 4 and 8 and removal strands $\bar{2}$ and $\bar{6}$, toehold-mediated strand displacement reactions take place. Rows 4 and 8 are blocked and the AuNCs are released from these rows. Meanwhile, rows 2 and 6 are activated to bind the AuNCs. As a result, the two AuNCs slide the filaments relative to one another for one step in a cooperative manner, introducing a 14 nm displacement.

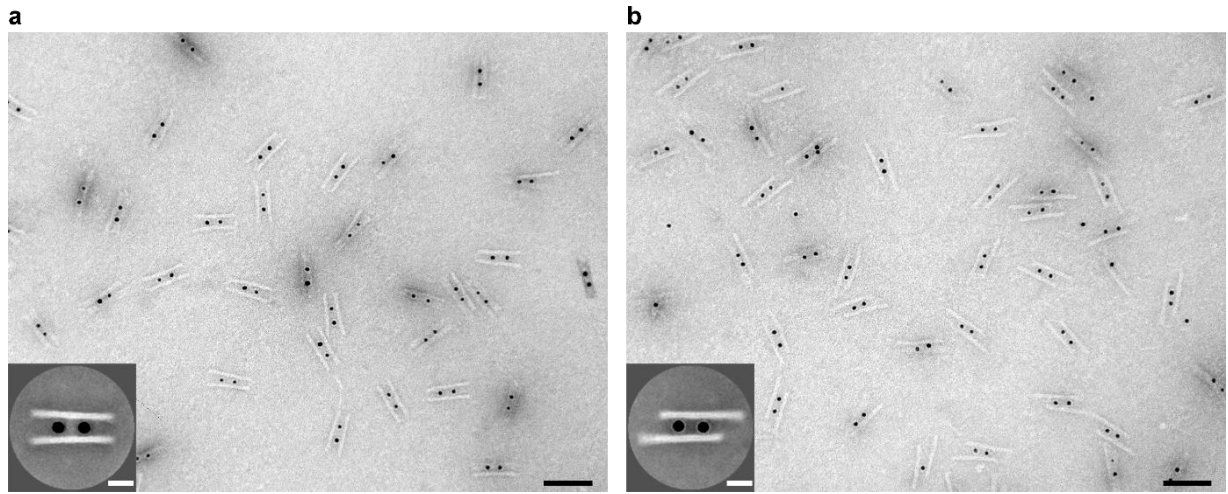

**Figure 2 | TEM images of the AuNC-origami structures. a,** TEM image of the structures before sliding. In the individual structures, two AuNCs are assembled in between two DNA origami filaments. Scale bar, 100 nm. Inset: averaged TEM image. Scale bar, 20 nm. (**b**) TEM image of the structures after two sliding steps. Scale bar, 100nm. Inset: averaged TEM image. Scale bar, 20 nm.

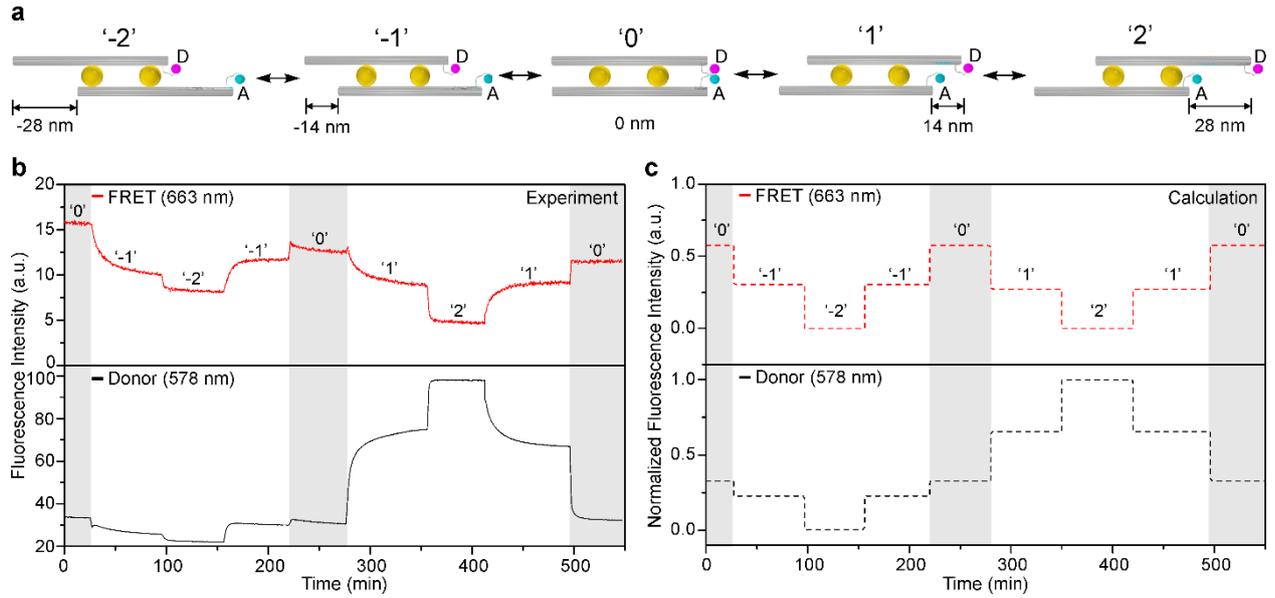

**Figure 3 | AuNC-mediated relative sliding monitored by *in situ* fluorescence spectroscopy. a**, Schematic of the five sliding states '-2', '-1', '0', '1', and '2'. D and A represent donor and acceptor, respectively. **b**, Experimental FRET and donor fluorescence signals monitored at wavelengths of 663 nm and 578 nm, respectively, with excitation wavelength of 530 nm. **c**, Theoretical calculation including both FRET between the fluorophores and electromagnetic quenching of the fluorophores by the AuNCs.

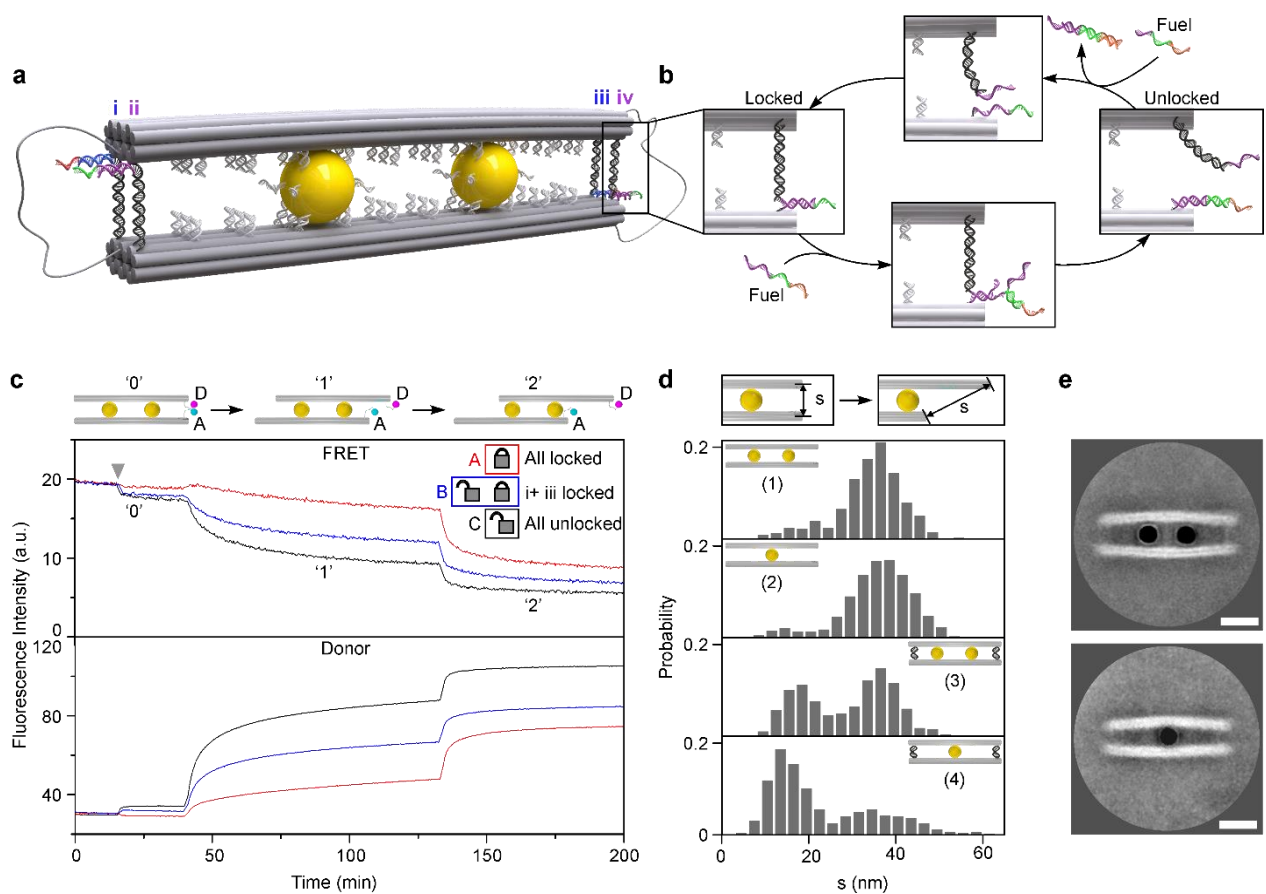

**Figure 4 | Relative sliding in the presence of DNA side-locks. a**, Schematic of the four DNA side-locks (i, ii, iii, and iv) introduced at the two ends of the DNA origami structure. Two differently sequenced segments (in blue and purple) are used to selectively lock and unlock the locks. **b**, Schematic of the locking and unlocking mechanisms of the DNA side-locks. **c**, FRET and donor fluorescence signals monitored during sliding for samples with four (A), two (B), and zero (C) locks. The grey arrow indicates the starting position of the experiments. **d**, Histograms of the end-to-end distance (s) of the filaments after two sliding steps in dependence on the DNA side-locks and on the AuNC number. **e**, Averaged TEM images for the locked structures with one and two AuNCs, respectively. Scale bars: 20 nm.